\documentclass[prl,twocolumn,showpacs]{revtex4}
\usepackage{graphics}
\begin{document}
\title{Inversion of magnetoresistance 
in magnetic tunnel junctions : effect of pinhole nanocontacts}
\author{Soumik Mukhopadhyay}
\author{I. Das}
\affiliation{ECMP Division, Saha Institute of Nuclear Physics, 1/AF,Bidhannagar, Kolkata 700064, India}
\begin{abstract}
Inverse magnetoresistance has been observed in magnetic tunnel
junctions with pinhole nanocontacts over a broad temperature
range. The tunnel magnetoresistance undergoes a change of sign at
higher bias and temperature. This phenomenon is attributed to the
competition between the spin conserved ballistic transport through the 
pinhole contact where
the transmission probability is close to unity and spin polarized
tunneling across the insulating spacer with weak transmittivity.
\end{abstract}
\maketitle
\noindent
During the last decade, study of spin polarized tunneling in 
Magnetic Tunnel Junctions (MTJ)~\cite{jag} has experienced an exponential
growth. Recently, the theoretical prediction and subsequent observation of 
large room temperature magnetoresistance
in epitaxial Fe/MgO/Fe structure~\cite{mgo} 
have generated tremendous interest among the physicists 
both from fundamental and technological point of view.   
However, the influence of ballistic spin dependent transport
(due to the presence of pinhole nanocontacts which connect the two 
ferromagnetic electrodes) on the magnetoresistive properties of MTJs 
has not been explored substantially. 
Recent simulations have shown that as much as $88\%$ of the current can
flow through the pinholes~\cite{simulation} in MTJs even though the 
bias dependence of differential
conductance has positive curvature. There are reports of
large Ballistic Magnetoresistance (BMR) at room temperature~\cite{bmr1} 
in ferromagnetic
nanocontacts. The essential ingredient for BMR is the condition of
nonadiabaticity in ballistic transport across the nanocontact. If the
domain wall width at the nanocontact is sufficiently thin so that the spin does
not have time to flip then the situation becomes analogous to the spin
conserved tunneling in MTJs~\cite{bmr}. In that case the BMR is related to the
spin polarization of the electrodes in the same way as the Tunneling
Magnetoresistance (TMR) in Julliere~\cite{julliere} or Slonczewski's 
model~\cite{slonc} which predicts positive TMR for symmetric electrode MTJ.
However, it is claimed~\cite{ballistic1} that ballistic channels in MTJs 
are not magnetoresistive and the opening up of a spin-independent 
conduction channel can only reduce the TMR.
We will show that the ballistic channel in MTJs are not only magnetoresistive,
it, in fact, can cause inverse tunneling magnetoresistance.
The relative contributions from the two
conduction channels -- elastic tunneling through the insulating spacer
and ballistic spin polarized transport through the narrow pinhole shorts -- can
change as the temperature and applied bias are varied and
magnetoresistive response can change accordingly.

\indent
Observation of inverse tunneling magnetoresistance (TMR) where the conductance 
in the antiparallel magnetic configuration is higher than that in the
parallel configuration, has been instrumental in understanding some of the 
important aspects of spin
polarized transport in MTJs. For example, the inverse TMR observed in 
experiments by De Teresa et. al.~\cite{desci}
have proved that the transport properties of MTJ
depend not only on the ferromagnetic metal electrodes but also on the 
insulator. Generally inverse TMR can occur if
the sign of spin polarization of the two electrodes is opposite in the
relevant bias range. 

\indent
An interesting inversion of TMR has been observed by us in a broad temperature 
range. The observed inverse TMR (TMR$=\delta R/R = (R_{AP}-R_{P})/R_{P}$ 
where $R_{AP}$, $R_{P}$ are the junction resistances
in antiparallel and parallel magnetic configuration of the MTJs respectively.) 
changes sign as the bias voltage and temperature is increased. The inverse 
TMR is attributed to the spin conserved transport through nanoscale metallic
channel where the transmission probability is close to unity.
Spin polarized tunneling with weak transmission probability 
dominates at higher temperature and bias leading to normal positive TMR.

\indent
The trilayer La$_{0.67}$Sr$_{0.33}$MnO$_{3}$
 (LSMO) / Ba$_{2}$LaNbO$_{6}$ (BLNO) / LSMO
was deposited on single crystalline SrTiO$_{3}$ (100) substrate
held at a temperature $800^{0}$C and oxygen pressure $400$ mTorr,
using pulsed laser deposition. Thickness
of the bottom LSMO layer is $1000\AA$  
and that of the top layer $500\AA$ while the
estimated thickness of the insulating spacer from the deposition
rate calibration of BLNO is $50\AA$. The microfabrication 
was done using photo-lithography and ion-beam milling. For details
see ref.~\cite{soumik}.

\begin{figure}
\resizebox{4cm}{3.2cm}
{\includegraphics{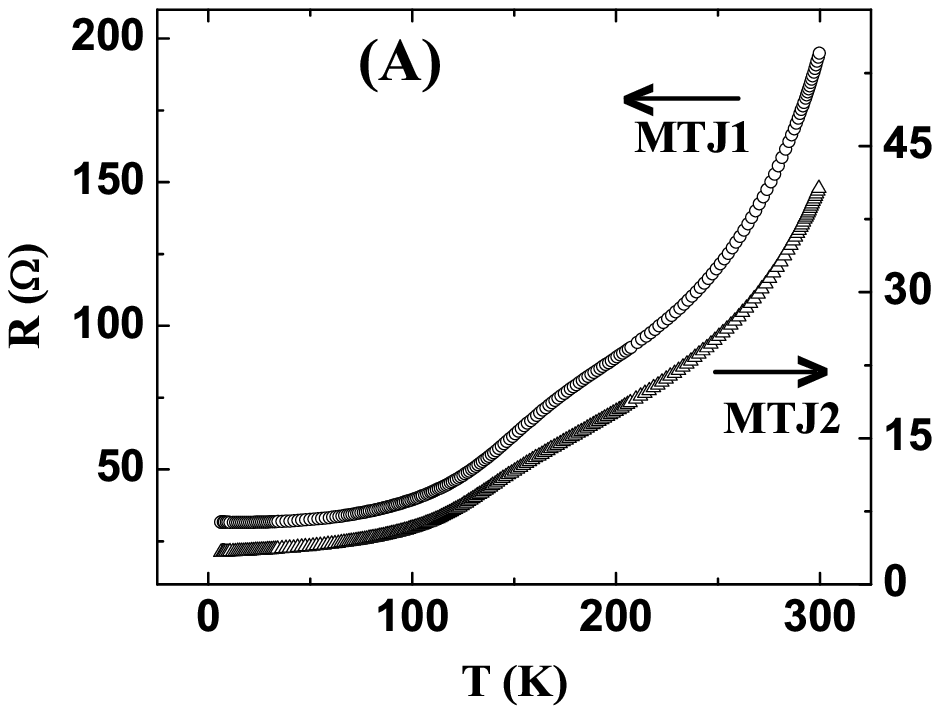}}
\resizebox{4.5cm}{3.4cm}
{\includegraphics{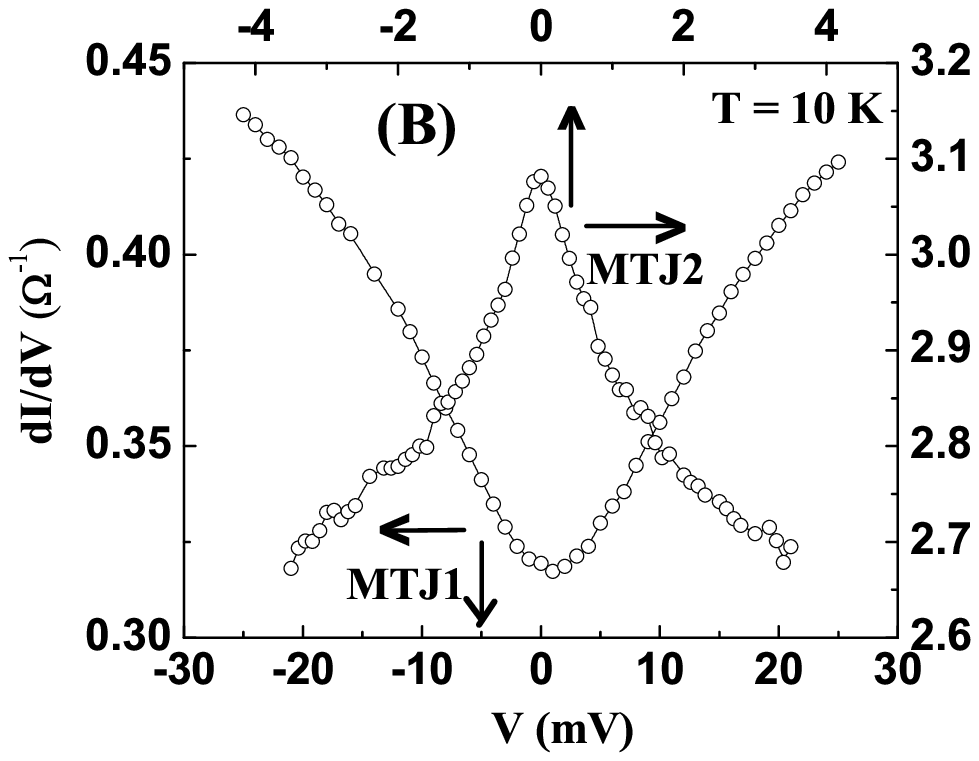}}
\caption{(A) Junction resistance vs. temperature curves for pinhole-short
MTJ1 and MTJ2 showing metal-like temperature dependence of resistance.
(B) Non-ohmic voltage dependence of differential conductance showing
opposite curvatures for MTJ1 and MTJ2.}\label{fig:cond}
\end{figure}

\indent
There is a set of criteria, known 
as Rowell's criteria~\cite{rowell}, 
for determining the quality of the tunnel junction.
However, for magnetic tunnel junctions, only three of these criteria
are applicable. $1)$ Exponential thickness dependence of junction resistance.
$2)$ Parabolic differential conductance curves that should be well fitted
by rectangular barrier Simmons~\cite{sim} model or trapezoidal barrier Brinkman
model~\cite{bri}. $3)$ Insulating like temperature
dependence of junction resistance. It has been observed that MTJs with
pinhole shorts can reproduce the first two criteria~\cite{short1}. 
Therefore the third
criteria stands out as the reliable proof of the quality of the MTJ.
Although the junctions show non-ohmic voltage dependence, 
the temperature dependence of junction resistance is metal-like
(Fig:~\ref{fig:cond}A). In this letter, we will show that two MTJs with 
pinhole shorts exhibit almost identical magnetoresistive properties 
although the voltage dependence of differential conductance curves 
have opposite curvatures. 
While the sample denoted MTJ1 shows positive curvature in the conductance
curve, the conductance of MTJ2 has negative curvature (Fig:~\ref{fig:cond}B). 
The MTJs contain metallic nanocontacts through which electrons travel 
ballistically at low temperature and bias. However, at higher bias, 
``hot electron'' 
transport through the pinholes results in heat dissipation within the 
nanocontact region just outside the ballistic channel~\cite{heat} and  
thus increasing the resistance. At higher bias the back-scattering
into the narrow channel increases due to larger phonon density
of states at the nanocontact, which reduces the transmittivity 
resulting in negative curvature in the voltage 
dependence of differential conductance. However the conduction
channel due to tunneling will become less resistive at higher bias since then
the electrons will tunnel across relatively thin trapezoidal part
of the barrier. As a result, the pinhole short will produce 
negative curvature in
the differential conductance curve while tunneling should cause positive
curvature. Although transport in both the MTJs is dominated by conduction
through pinhole shorts which is evident in Fig:~\ref{fig:cond}A, the strong
positive curvature in the voltage dependence of conductance due to tunneling
can overcome the weak negative curvature due to transport through the pinholes,
resulting in overall positive curvature as observed in MTJ1 
(Fig:~\ref{fig:cond}B). Fitting the differential conductance curves
with positive curvature for MTJ1 by Brinkman model, 
the extracted barrier height turns out to be 
about $0.8-1$ eV (much higher than the value $0.2-0.3$ eV corresponding
to MTJs without pinhole shorts) and the barrier width much smaller $15-20\AA$
compared to that of $\sim40\AA$ for good MTJs. The extracted value
for barrier height increases 
while the barrier width decreases as the temperature is increased.
Although the value of the barrier parameters, in the present case, 
carry no physical significance,
temperature dependence of the barrier parameters is a reconfirmation of
the MTJ having pinhole shorts~\cite{short}. 

\begin{figure}
\resizebox{6cm}{6.5cm}
{\includegraphics{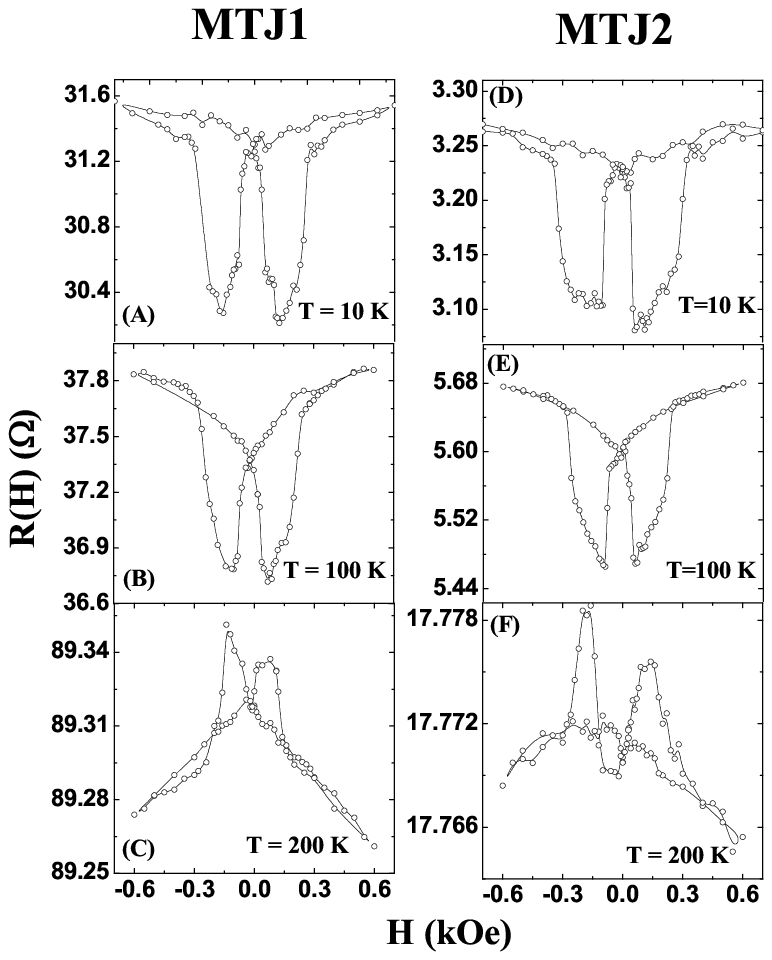}}
\caption{
Junction resistance vs. magnetic field curves for pinhole-short
MTJ1 (A, B, C) and MTJ2 (D, E, F) at different temperatures
showing the inverse TMR at low temperature which undergoes a change of sign as
temperature is increased to $200$ K.}\label{fig:mrht1}
\end{figure}

\begin{figure}
\resizebox{4cm}{4cm}
{\includegraphics{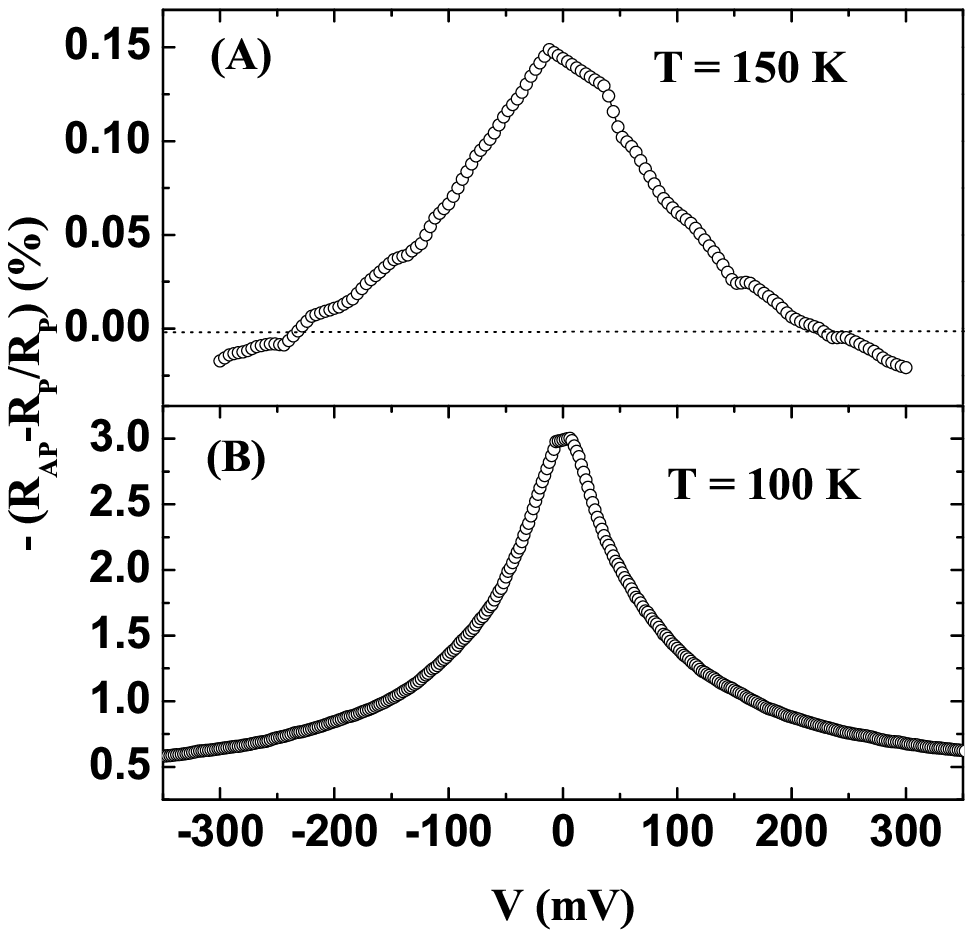}}
\resizebox{4cm}{4cm}
{\includegraphics{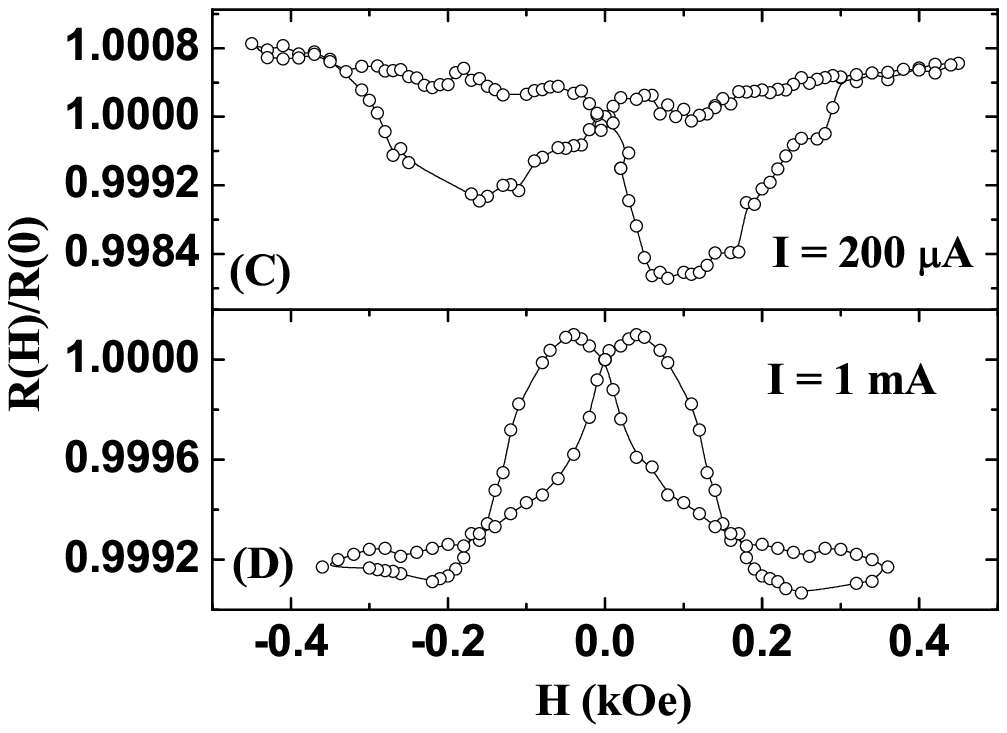}}
\caption{(A) Bias dependence of TMR for MTJ1 at $150$ K showing
bias induced inversion of magnetoresistance above $\pm 225$ mV.
(B) Bias dependence of TMR for MTJ1 at $100$ K showing no evidence
of sign change of TMR.
Below, The reduced junction resistance
vs. magnetic field curves for MTJ1 at $150$ K at bias currents $I = 200 \mu$A
(C) and $I = 1$ mA (D). At low bias, inverse TMR is observed while at
high bias the sign of TMR reverses resulting in positive TMR.}
\label{fig:mrbias1}
\end{figure}

\indent
Inverse TMR is observed for both MTJs over a broad temperature range 
$10-150$ K (Fig:~\ref{fig:mrht1}). The value of inverse TMR decreases 
with increasing temperature. For MTJ1 the value of inverse TMR 
is $4.6\%$ at $10$ K which reduces to about $1.8\%$ at $150$ K while for 
MTJ2 it is about $6.5\%$ at $10$ K, which almost vanishes at $150$ K.
Above $150$ K, the situation is the opposite -- ordinary positive 
TMR is observed. At $200$ K, the positive TMR exhibited 
by MTJ1 is about $1\%$ while that for MTJ2 is $0.06\%$ 
(Fig:~\ref{fig:mrht1}C,F). 
The bias dependence of TMR for MTJ1
has some interesting features. At $150$ K, it is observed that 
above $\pm 225$ mV, the TMR changes sign (Fig:~\ref{fig:mrbias1}A). 
A clear evidence of such inversion is highlighted in  
Fig:~\ref{fig:mrbias1}C,D 
where MTJ1 shows inverse TMR at bias current $I = 200 \mu A$
while at $I = 1 mA$, exhibits positive TMR. 
However, at a lower temperature $100$ K,
there is no evidence of such inversion with increasing bias 
(Fig:~\ref{fig:mrbias1}B).  

\indent
The observed phenomenon can be explained as follows. The present system can be
considered as being equivalent to two ferromagnetic metal electrodes
connected by ballistic nanoscale metallic channels
along with a conduction channel connected in parallel which describes
tunneling across the insulating spacer.
For the case of two identical ferromagnets connected by a
nanocontact, the ballistic magnetoresistance (BMR)~\cite{bmr} is given by,
\[\Delta R/R_{P}=\frac{2P^{2}}{1-P^{2}}f(k_{F}\lambda)\]
where $P$ is the spin polarization, 
$\lambda$ is the domain wall width and $k_{F}$ is the Fermi wave
vector, $f$ being the measure of the spin non-conservation in the current
through the nanocontact. In the limit of vanishing domain wall width $\lambda$,
spin flipping by domain wall scattering is absent. Then $f$ is unity
and the electron spin is conserved during transmission (the factor $f$ 
decreases with the increase of the product $k_{F}\lambda$). Hence we
arrive at the well known Julliere formula for tunneling magnetoresistance.
Thus there seems to be no difference in the spin conserved ballistic transport
through nano-sized pinholes or elastic spin polarized tunneling.
However there is a stark contrast in the transmittivities for the two
conduction channels.
In case of normal elastic tunneling through insulating barrier 
the tunneling probability is finite
but small and decays exponentially with increasing barrier width. 
On the other hand, the transmittivity through the metallic pinhole 
nanocontact is close to unity.

\indent
At low temperature,
the electron transfer from one ferromagnetic lead to another occurs
dominantly through the metallic pinhole shorts. Hence, according to 
ref.~\cite{bmr}, the ballistic magnetoresistance should follow
the Julliere or Slonczewski's model for spin polarized 
tunneling and should give positive
TMR for MTJs with identical electrodes. However, in our case, inverse
TMR is observed. The reason probably lies in the fact that the model
does not take into account the effect of high transmittivity and the 
possibility of different transmission coefficients
of the electrons in the majority and minority spin bands.
The model reduces to Julliere model in the non-adiabatic limit. 
However, there is a general agreement that the generalized Julliere model
is valid in the limit of very weak transmission probability~\cite{zhang}.
Tae-Suk Kim~\cite{kim} has very recently put forward 
a theoretical model for spin polarized transport
through a narrow channel. In this treatment,
when the spin is conserved in transport through a nanoscale channel
and the transmittivity is close to unity, there is a possibility
of inverse TMR. According to Kim's model, 
transmission probabilities in the parallel and anti-parallel
magnetic configuration of the two electrodes
(assuming that the spin polarizations of the two electrodes 
are the same) are given as,
\begin{eqnarray*}
T_{P} & = & \frac{2{\gamma}_{+}}{(1+{\gamma}_{+})^{2}}+\frac{2{\gamma}_{-}}{(1+{\gamma}_{-})^{2}}\\
T_{AP} & = & \frac{4\sqrt{{\gamma}_{+}{\gamma}_{-}}}{(1+\sqrt{{\gamma}_{+}{\gamma}_{-}})^{2}}
\end{eqnarray*}
where $\gamma_{+}$ and 
$\gamma_{-}$ are the transfer rates for majority and minority spins 
respectively. 
When the transmission probability is small, i.e. $\gamma_{\pm}<<1$,
$T_{P}=2(\gamma_{+}+\gamma_{-})$, $T_{AP}=4\sqrt{\gamma_{+}\gamma_{-}}$
which means that the transmission probability in the parallel configuration
is greater than that in the anti-parallel configuration i.e. the TMR is
positive.
The conditions for zero TMR or $T_{P}=T_{AP}$ are given as,
${\gamma_{+}}-{\gamma_{-}}=0$
which is a trivial solution and implies that spin polarizations at the Fermi
level for both the electrodes is zero and is applicable for nonmagnetic 
tunnel junctions. The non-trivial solution for zero TMR with spin polarization 
$P\neq0$, resides at the boundary
between two regions corresponding to $T_{P}>T_{AP}$ and $T_{P}<T_{AP}$
and is given by,
\[({\gamma_{+}}{\gamma_{-}}-1)^{2}-2\sqrt{{\gamma_{+}}{\gamma_{-}}}(1+\gamma_{+})(1+\gamma_{-})=0\]
To be more precise, the combinations $({\gamma_{+}},{\gamma_{-}})$ satisfying
the above equation constitutes a curve in $({\gamma_{+}},{\gamma_{-}})$ space 
enclosing the region where normal positive TMR occurs. The region outside
the curve contains high values for ${\gamma_{+}}$ and ${\gamma_{-}}$
which corresponds to inverse TMR.
The transmission probabilities for the majority and minority
spin band are related to $\gamma_{\pm}$ as follows,
\[T_{\pm}=\frac{4\gamma_{\pm}}{{(1+\gamma_{\pm})}^{2}}\]
Thus, when transmission probability is closer to unity i.e. 
$T_{\pm}\simeq 1$ and there
is an imbalance in the transmission probabilities for the majority spin
and the minority spin, inverse TMR occurs. 
\begin{figure}
\resizebox{6.5cm}{5.5cm}
{\includegraphics{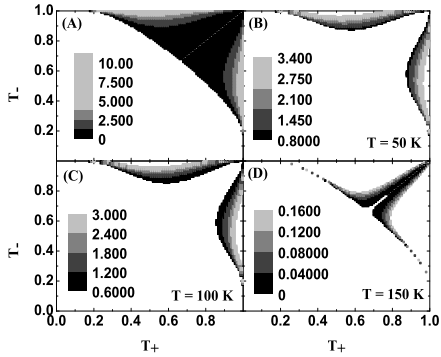}}
\caption{{\bf A:} The theoretically allowed values of $T_{\pm}$ for
inverse TMR in the $\{T_{+},T_{-}\}$ plane and the corresponding values
of inverse TMR is shown by a color map. {\bf B,C,D:} The
allowed values of $T_{\pm}$ which causes inverse TMR for MTJ1 at
$50$, $100$, $150$ K respectively within the bias range $\pm 220$ mV.
The corresponding values of inverse TMR in each case is shown by
the color map.}\label{fig:phase}
\end{figure}

\indent
Replacing $\gamma_{\pm}$ by $T_{\pm}$ in the expression for
$T_{P}$ and $T_{AP}$, the TMR values ($\Delta R/R_{P}=\{T_{P}-T_{AP}\}/T_{AP}$)
can be calculated for all possible values of $T_{\pm}$.
The theoretically allowed values of ($T_{+}$, $T_{-}$) for
inverse TMR and
how the allowed values of ($T_{+}$, $T_{-}$)
evolve with the change in temperature for MTJ1, within the bias range 
$\pm 220$ mV, are shown in Fig:~\ref{fig:phase} along with the
corresponding TMR values. The contribution due to the parallel
tunneling conduction channel has been neglected for simplicity of
calculation. This will, of course, lead to underestimation of the allowed
values of $T_{\pm}$ particularly in the high temperature region
where the relative contribution of the tunneling conduction channel will
be substantial.
The calculation suggests that, larger the imbalance
between $T_{+}$ and $T_{-}$, the greater is the value of
inverse TMR as can be seen from Fig:~\ref{fig:phase}(A). 
Up to $100$ K,
the allowed values of $T_{\pm}$ stay away from the 
$T_{+}=T_{-}$ line (Fig:~\ref{fig:phase};B,C). 
However, as the temperature is
increased further, the imbalance in the transfer rates of majority
and minority spins diminishes drastically and the allowed values
congregate near $T_{+}=T_{-}$ 
(Fig:~\ref{fig:phase}D). 
The increase in bias also reduces the imbalance between the transmittivities
in the two bands as can be seen from the color map for each temperature.
If the values of 
$T_{+}$ and $T_{-}$ are interchanged the TMR
remains the same. However, the physically acceptable situation
is where $T_{+}$ is greater than $T_{-}$, since the minority
spin states are generally regarded as being more localized
compared to the majority spin states. 

\indent 
Although La$_{0.67}$Sr$_{0.33}$MnO$_{3}$ is generally considered to be
having almost full spin polarization,
Andreev reflection experiments have confirmed the existence of minority spin 
states which will be particularly 
influential in the ballistic limit of transport~\cite{minority}.  
The change from inverse TMR to a positive one at higher bias at $150$ K
can be attributed to the fact that at higher
bias electrons tunnel through relatively thin trapezoidal part of
the barrier such that the contribution due to elastic tunneling increases which
gives rise to positive TMR. On the other hand, there are several reasons
for the decrease of inverse TMR at higher bias due to transport through 
the narrow channel.
$1)$ Local generation of heat within the nanocontact region at higher bias
leads to increased thermal spin fluctuation and resistance 
at the nanocontact which reduces the inverse TMR. $2)$ The back-scattering
into the narrow channel increases as a result of larger phonon density
of states at the nanocontact, reducing the transmittivity and hence the
inverse TMR.  
$3)$ Lastly, the product $k_{F}\lambda$ being larger at higher
bias may cause deviation from the non-adiabatic limit. This results in increased
spin flip scattering thus reducing the magnitude of inverse TMR.
In our case, the normal positive
TMR is observed at $200$ K, where elastic tunneling across the
insulating spacer with weak tunneling probability is dominant 
and the electron-phonon interaction is large
enough to push the transport through the pinhole into diffusive regime.
Manganese Oxide tunnel junctions
with pinhole shorts are better suited to exhibiting inverse TMR than MTJs
with transition metal electrodes since in that case 
there is high probability of the
pinhole shorts getting oxidized which would lead to weak transmittivity
through the narrow channel.

\indent
To summarize, we have presented a direct experimental evidence that pinhole
shorts through the insulating spacer in a magnetic tunnel junction
can cause inverse tunneling magnetoresistance when the transmission
probability is close to unity, which is an indicator that Julliere
and Slonczewski models are no longer valid in this regime. 
The relative
contributions from the conduction channels due to elastic tunneling and
ballistic spin conserved transport through the pinholes can be changed
by proper adjustment of the bias and temperature, which can even result in
the change of sign of the tunneling magnetoresistance.

\indent
The authors acknowledge Mr. S. P. Pai for his technical help
during micro-fabrication of the MTJs.

\end{document}